\documentstyle[aps,multicol]{revtex}
\begin{document}

\title{Non locality, closing  the detection loophole 
and communication complexity}

\author{Serge Massar}             
\address{
Service de Physique Th\'eorique,
Universit\'e Libre de Bruxelles, CP 225, Bvd. du Triomphe, B1050
Bruxelles, Belgium.
}
\date{17-08-01}

\maketitle

\begin{abstract}
It is shown that the detection loophole which arises when trying to
rule out local realistic theories as alternatives 
for quantum mechanics 
can be closed if the detection 
efficiency $\eta$ is larger than $\eta \geq d^{1/2}
2^{-0.0035d}$ where $d$ is the dimension of the entangled system. 
Furthermore it is argued that this 
exponential decrease of the detector efficiency required to close the
detection loophole is almost optimal. This argument is based on a
close connection that exists between  closing the
detection loophole and the
amount of classical 
communication required to simulate quantum correlation when the
detectors are perfect.
\end{abstract}

\vspace{0.2cm}

\begin{multicols}{2}

%\section{introduction}

Experimental tests of the entanglement of quantum systems are 
important for several reasons.
They provide an experimental check of the
validity of quantum mechanics, and in particular the surprising ``non
locality''  
exhibited by quantum mechanics.
Furthermore they can be viewed as primitives from
which one can build more complicated protocols of interest for
quantum information processing and they provide a benchmark with which to 
compare the performance of different quantum systems, such as ion
traps, photons, etc. 

To test the entanglement of a quantum system one carries out
measurements on each particle, and compares the correlations between
the results of these measurements with the predictions of quantum
mechanics. A crucial check of the quantumness of these correlations is
whether they exhibit ``non locality'', that is whether they 
cannot be reproduced by a classical local variable
theory (also called local realistic theory)\cite{Bell}. 
Formally this is done by inserting the joint probabilities
of outcomes into an
inequality, called a ``Bell inequality'', 
which must be satisfied in the case of local variable
theories but can be violated by quantum mechanics.

During the past decades successively more
sophisticated tests of Bell inequalities have been carried out (for a
review see\cite{A}). 
Most experiments so far have
involved entangled photons. By letting the photons propagate 
a large distance from their emission point it has been possible to
spatially separate the two measurements and thereby close the so
called ``locality loophole''. However in optical experiments, because
of losses and small detector efficiency, all tests of Bell
inequalities so far leave open the so called 
``detection loophole''. This means that all experimental results 
that use
pairs of photons  
can be explained by a classical local variable theory 
if the local
variable theory can instruct the detectors either to click,
i.e. register the presence of a particle, or not.
The strongest theoretical result so far is that the detection loophole
can be closed in the efficiency is $\eta > 2/3$\cite{E}, 
but this is too stringent for optical experiments.
Recently an experiment that closes the
detection loophole has been carried out using trapped ions\cite{ions}. 
But in this experiment the ions where separated by a very small
distance and the locality loophole was not closed.

In almost all experiments on entangled systems 
each system belongs to a  Hilbert space of dimension  2. 
(One recent experiment
tested the entanglement of systems of dimension 3 \cite{B}). 
However when pairs of photons are produced (for instance by
parametric down conversion), the photons
are entangled in position-momentum and time-energy in addition to a
possible entanglement in polarization. Thus entangled systems of 
large dimensionality can easily be produced in the laboratory. Can one 
exploit the large dimensionality of these entangled photons to
carry out stronger tests of quantum non locality? This has been
the subject of several recent theoretical works \cite{1,2,3,4,us} in
which 
it has 
been shown that using entangled
systems of large dimensionality can be advantageous, but no spectacular
improvements have been found. 

In the present work it will be shown that using entangled systems of
large dimensionality allows in principle a dramatic decrease in the
detector efficiency required to close the detection loophole. More
precisely, the minimum detector efficiency required
to close the detection loophole decreases exponentially with the
dimension $d$.
This is particularly relevant to possible
experiments involving momentum or energy entangled photons since 
in this case it may be possible to devise an experiment in which  
photon losses and detector efficiency
decrease only slowly with the dimension.

This result is obtained by explicitly describing a set of measurements 
carried out by Alice and Bob on an entangled system of large dimension
and writing a Bell inequality adapted to this measurement scenario. It 
will be shown that this Bell inequality is violated even for
exponentially small detector efficiencies. However this Bell
inequality is extremely sensitive to noise and therefore does not
constitute a realistic experimental proposal.
A noteworthy feature of this measurement scenario is that the number
of measurements between which Alice and Bob must choose is
exponentially large. 

In the second part of this letter we consider whether it is possible
to improve this Bell inequality. Can one  decrease the number of
measurements between which Alice and Bob must choose, or decrease 
the dimensionality of the entangled system, while keeping the same low 
sensitivity to detector inefficiency? We argue that this is not the
case and that our Bell
inequality is close to optimal. 

These latter results 
follow from a close connection between
the detection loophole and the minimum amount of classical communication
required to perfectly simulate measurements on an entangled quantum system.
Suppose measurements are carried out on an entangled
 quantum system (with perfect detectors $\eta =1$). The correlations
 exhibited by such measurements will in general violate a Bell inequality and
 therefore cannot be reproduced by local variable theories. However by 
 supplementing the local variable theory by classical communication one 
 can reproduce the quantum correlations. Recently there have been
 several works that attempted to understand how much classical
 communication is necessary to bridge the gap between quantum
 mechanics and local variable theories\cite{BCT,S,MBCC}.
Intuitively one would expect that the more communication is required
to recover the quantum correlations, the stronger the quantum
correlations test non locality. This intuition will be made precise
below in the context of the detection loophole. 
It will be shown that 
the minimum amount of classical communication $C^{min}$ required to
recover the quantum correlations is anti-correlated to 
the minimum detection
efficiency $\eta^*$ required to close the detection loophole.

We begin with some definitions.

{\em A measurement scenario} is defined by a bipartite quantum state
$\psi$ belonging to the tensor product of two Hilbert spaces $H_A
\otimes H_B$, and by two ensembles of measurements, $M_A$ acting on $H_A$
and 
$M_B$ acting on $H_B$.
For instance $\psi = \sum_{k=1}^d |k\rangle_A|k\rangle_B / \sqrt{d}$ 
can be the maximally entangled state of $d$
dimensions. The elements $x \in M_A$ are a basis of $H_A$: 
$x=\{|x_1\rangle , ...,
|x_d\rangle \}$ with $\langle x_i | x_j \rangle =
\delta_{ij}$. Similarly the elements $y \in M_B$ are a basis of $H_B$.
Party A is given as input a random element $x \in M_A$ and party B 
is given as input a random element $y \in M_B$. 

In a measurement scenario with perfect detectors ($\eta =1$), both
parties must give as output
one of $d$ possible outcomes. Denote Alice's output by $a$ and
Bob's output by $b$. The joint probabilities of the outcomes are
$P(a=i,b=j|x,y) =  |  \langle \psi|x_i\rangle |y_j\rangle |^2$.

In a measurement scenario with detectors of finite efficiency $\eta$, both
parties must give as output
one of $d+1$ possible outcomes.
Output $0$ occurs with probability $1-\eta$ and corresponds to the detector
not detecting the particle whereas outcomes $1$ to $d$ occur with
probability $\eta$ and correspond to
a specific result of the measurement when the particle is
detected. The
probability that one of the detectors gives outcome $0$ is independent
of the other detector. Thus
the joint probabilities of outcomes are:
\begin{eqnarray}
P(a=0,b=0|x,y) &=& (1-\eta)^2\ ,\nonumber\\
P(a=i,b=0|x,y) &=& \eta (1-\eta)\mbox{Tr} |x_i\rangle\langle x_i|
\otimes \openone_B 
|\psi\rangle\langle \psi|\ , \nonumber\\
P(a=0,b=j|x,y) &=& 
\eta (1-\eta) \mbox{Tr}\openone_A \otimes  |y_j\rangle\langle y_j| 
|\psi\rangle\langle \psi|\ ,\nonumber\\
P(a=i,b=j|x,y) &=& \eta^2 |  \langle \psi|x_i\rangle |y_j\rangle |^2
\ .
\label{joint}
\end{eqnarray}

{\em In a local variable theory for the measurement scenario $\{\psi,
M_A, M_B\}$ with detector efficiency $\eta$}, 
Alice and Bob are both given the same 
element $\lambda \in 
\Lambda$ drawn with probability $p(\lambda)$ (often called the ``local 
hidden variable''). Alice knows $x$ but does not know $y$. From her
knowledge of $\lambda$ and $x$, Alice selects an outcome
$a=f(x,\lambda)$. 
Similarly Bob
knows $y$ but does not know $x$ and chooses an outcome 
$b=g(y,\lambda)$. We can suppose that the functions $f$ and $g$ are
deterministic since all local randomness can be put in $\lambda$.
The joint probabilities $P(a,b|x,y) = 
\int_\Lambda d\lambda \ p( \lambda) \delta (f(x,\lambda) -a)
\delta( g(y,\lambda) -b)$ 
are identical with the predictions of quantum mechanics eq. (\ref{joint}).

A local variable theory will only exist if the detector efficiency is
sufficiently small. 
The maximum detector efficiency for which a local variable
theory exists will be denoted 
$\eta^*(\psi, M_A, M_B)$.

We are now in a position to state our main result:

{\em Theorem 1:} There exists a measurement scenario for which the
state is the maximally entangled state of dimension $d=2^n$ with
$n\geq 2$
an integer, and for which the number of measurements carried out by
Alice and Bob are exponentially large $|M_A|=|M_B| = 2^d$, and such that
the detection
loophole is closed if  $\eta \geq d^{1/2} 2^{-0.0035 d}$.

{\em Proof:}
We consider the same measurement scenario as that 
described in Theorem 4 of \cite{BCT} (which is 
inspired by the Deutch-Jozsa
problem,
see \cite{BCW}).
The state is $\psi =
\sum_{k=1}^{d=2^n}|k\rangle|k\rangle/\sqrt{d}$. 
The sets of measurements $M_A$ and $M_B$ are identical. 
The  measurements  $x\in M_A$ 
are parameterized by a string of $d$ bits: $x=x_1x_2...x_{d}$ where $x_i
\in\{0,1\}$ and similarly for $y \in M_B$. Hence $|M_A|=|M_B| = 2^d$. 
The measurements are
described in detail in \cite{BCT}. They have the important properties that 

1. if $x=y$, then Alice and Bob's outcome are
  identical ($a=b$),

2. if the Hamming distance $\Delta(x,y)$ between $x$ and $y$ is $
\Delta(x,y) = d/2$, then
  Alice and Bob's outcomes are always different ($a \neq b$).

Let us define $\alpha(x,y) =  \delta(x=y) - \delta(\Delta(x,y)=d/2)$ 
which is equal to $+1$ if $x=y$, equal to $-1$ if
$\Delta(x,y) = d/2$, and equals zero otherwise.
Consider the following Bell expression
\begin{equation}
I = \sum_{x=1}^{2^d} \sum_{y=1}^{2^d}
P(a=b \mbox{ AND } a \neq 0 ) \alpha(x=y) \ .
\label{Ibell}
\end{equation}
It is immediate to compute the value of $I$ predicted by quantum
mechanics for the above measurement scenario since from properties 1
and 2 above, only the term 
proportional to $  \delta(x=y)$ 
contributes:
\begin{equation}
I(QM) = \eta^2 2^d \ .
\label{IQM}
\end{equation}

It is more difficult to compute the maximum value of $I$ in the
case of local variable theories.
Let $Z$ be the largest subset of
$\{0,1\}^{d}$ such that if $z,z' \in Z$, then $\Delta(z,z') \neq d/2$ 
(i.e. no two elements of $Z$ are Hamming distance $d/2$ one from the
other). We shall show below that 
\begin{equation}
I(\mbox{local variable}) \leq d |Z| 
\label{IZ} \end{equation} 
independently of $\eta$.
Frankl and R\"odl have given bounds on $|Z|$. Theorem 1.10 of
\cite{FR} states that $|Z|<(2-\epsilon)^d$ for some constant
$\epsilon >0$. And from corollary 1.2 of \cite{FR} one can deduce a
more precise bound: $|Z| < 2^{.993}d$. 
Combining this with eq. (\ref{IQM}) implies that  
 one can close the
detection loophole if
$\eta   \geq   d^{1/2} 2^{-0.0035
  d}\geq   d^{1/2} |Z|^{1/2} 2^{-d/2}  $.

We now prove eq. (\ref{IZ}). Recall that in the case of
local variable model, Alice's output is a function $a(\lambda, x)$ of
the local variable and of her measurement, and similarly for
Bob. Using 
$P(a=b \mbox{ AND } a \neq 0 ) = \sum_{k=1}^d P(a=k  \mbox{ AND }
b=k)$, the value of $I$ for a local variable model can be written as
\begin{eqnarray}
I(\mbox{lv}) &=& 
\sum_\lambda p(\lambda) \sum_x \sum_y \sum_{k=1}^d \nonumber\\
& & \quad
 P[a(\lambda,x)=k  \mbox{ AND } b(\lambda,y)=k] \alpha(x,y)
\nonumber\\
&=& \sum_\lambda p(\lambda) \sum_{k=1}^d  
 \sum_{x \in X_{k\lambda}} \sum_{y\in Y_{k\lambda}} \alpha(x,y)
\label{IZZ}
\end{eqnarray}
where $X_{k\lambda}$ is the set of $x$ such that $a(\lambda,x)=k$ and
$Y_{k\lambda}$ is the set of $y$ such that $b(\lambda,y)=k$.
Let us denote by $Z_{k\lambda}$ the largest set such that 1)
$Z_{k\lambda} \subset X_{k\lambda}$; 2)  $Z_{k\lambda} \subset Y_{k\lambda}$;  
3) if $z,z' \in Z_{k\lambda}$ 
then $\Delta(z,z') \neq d/2$. This implies that 
$|Z_{k\lambda}| \leq |Z|$.
Consider the sum $\beta(x) = \sum_{y\in Y_{k\lambda}} \alpha(x,y)$.
$\beta(x)$ is an integer less or equal to $1$. Let us show that 
if $x\not\in Z_{k\lambda}$, then $\beta(x) \leq 0$. Suppose this is not
true (i.e. $x\not\in Z_{k\lambda}$ and $\beta(x) =1$), then necessarily
$x\in Y_{k\lambda}$ and there is no $y \in Y_{k\lambda}$ such that
$\Delta (x,y)=d/2$. But then we could increase $Z_{k \lambda}$ by
adding $x$ to $Z_{k \lambda}$. But $Z_{k \lambda}$ is maximal, hence
there is a contradiction.
We therefore obtain that
$\sum_{x\in X_{k\lambda}} \beta(x) \leq \sum_{x\in Z_{k\lambda}}
\beta(x) \leq |Z_{k \lambda}| \leq |Z| $. Inserting this in eq. (\ref{IZZ}) 
yields
eq. (\ref{IZ}).
$\Box$

Note that the Bell expression eq. (\ref{Ibell}) is extremely sensitive to 
noise. This is because in the presence of noise the term in $\alpha$
proportional to $\delta(\Delta(x,y)=d/2)$ receives a very large
contribution, and therefore leads to a much reduced value of $I$.

We now turn to the relation between the detection loophole and
communication complexity. We begin with a definition:

{\em  In a local variable theory supplemented by $C$ bits
of classical communication for the  measurement scenario $\{\psi,
M_A, M_B\}$ with perfect detectors ($\eta =1$)} the parties, 
in addition to sharing
the random variable $\lambda$, are allowed to communicate $C$ 
bits before choosing
there output. 
Note that  one should distinguish whether
 $C$ is the absolute bound on the amount of communication,
or whether $C$ is the average amount of communication between the
parties, where the average is taken over many repetitions of the
protocol, see\cite{MBCC}.

For a given measurement scenario $\{\psi,
M_A, M_B\}$ with perfect detectors 
one can try to minimize the amount of communication
required to reproduce the quantum probabilities.
The minimum amount of communication required to simulate the
measurement scenario in the average communication model will be denoted 
$ C^{min}(\psi,M_A, M_B)$.

We shall now show that 
the  minimum detector efficiency $\eta^*$ required to close the detection 
loophole 
and the minimum amount of communication $C^{min}$ 
required to simulate
a measurement scenario with perfect detectors are closely related.
We begin by showing  that if a measurement scenario is
difficult to simulate classically, then the minimum detector efficiency
required to close the detection loophole is small. In fact this result 
was the inspiration for Theorem 1: the measurement scenario considered 
in Theorem 1 is difficult to simulate
classically\cite{BCT}, hence $\eta^*$ must be
small. Further investigations led to the strong result of Theorem 1.

{\em Theorem 2:} For all measurement scenarios $\{ \psi, M_A,
M_B$\}, the relation $
 \eta^*  (\psi, M_A, M_B )
\leq \sqrt{ 2 /  C^{min}(\psi,M_A, M_B) }
$ holds.

{\em Proof}.
It will be shown that  any local variable model with
detector efficiency $\eta$ can be mapped into a communication protocol
with an average of $2 / \eta^2$ bits of communication.
Therefore $C^{min} \leq 2 / \eta^2$ for all detector efficiencies for
which a local variable model exists, and this yields the upper bound
on $\eta^*$.

Recall that a local variable model is defined by the two functions $f$ and
$g$ introduced above and the probability distribution $p$ on the
space $\Lambda$. Now suppose that initially the
parties share an infinite number of i.i.d. hidden variables
$\lambda_1,\lambda_2,\lambda_3,...$ each drawn from the space
$\Lambda$ with probability $p$. 
Consider the following protocol in which the two parties repeatedly 
simulate the
local variable model and communicate whether the model predicts
that the detectors work or not: 

1. Set the index $k=1$.

2. Alice computes $f(x,\lambda_k)$ and Bob computes $g(y,\lambda_k)$

3. Alice tells Bob whether $f(x,\lambda_k)=0$ or
$f(x,\lambda_k)\neq 0$ 
and Bob tells Alice
whether $g(y,\lambda_k)=0$ or $g(y,\lambda_k)\neq 0$.

4. If $f (x,\lambda_k)= 0$ or $g(y,\lambda_k) = 0$, Alice and Bob 
increase the index $k$ by 1 
and go back to step 2.

5. If $f(x,\lambda_k)\neq 0$ and $g (y,\lambda_k)\neq 0$ then Alice
outputs $f(x,\lambda_k)$ and Bob outputs $g(y,\lambda_k)$.

This protocol reproduces exactly 
the correlations exhibited by quantum mechanics.
The mean number of iterations of the protocol is $1/\eta^2$. The
number of bits communicated during each iteration is 2 (one bit from
Alice to Bob and one from Bob to Alice). Hence the average amount of
communication is $2/ \eta^2$. 
$\Box$

We now investigate whether a model with finite communication and perfect
detectors can be mapped into a local variable model with
inefficient detectors. We will give an argument, but not a proof, 
that suggests that
such a mapping should exist.

Consider a measurement scenario. 
Suppose there is a classical protocol that simulates the quantum
correlations with $C$ bits of
communication. In this protocol, Alice initially knows the local
variable $\lambda$ and her measurement $x$, and Bob initially knows
the local variable $\lambda$ and his measurement $y$. Denote the
conversation 
by ${\cal 
  C}(x,y,\lambda) = c_1 c_2 \ldots$ where $c_i \in \{0,1\}$ is the
i'th bit in the conversation. 
Alice and Bob's outputs are
therefore given by functions
$a=f(x,\lambda,{\cal C})$ and $b=g(y,\lambda,{\cal C})$.

Now suppose that in addition to the local variable $\lambda$, Alice
and Bob share a second local variable $\mu =\mu_1 \mu_2 \ldots$ which
consists of an infinite string of independent random  bits 
$\mu_i \in \{0,1\}$. The basic idea is that Alice and Bob will check
whether the local variable $\mu$ is a possible conversation 
$\mu ={\cal  C}(x,y,\lambda)$. If it is they give the corresponding
output. 
If it 
is not they give the outcome $0$ corresponding to the detectors not
working. The probability that $\mu ={\cal 
  C}$ is $2^{-C}$ which suggests that if $\eta \leq 2^{-C}$ a local
variable model should exist. 

Making the above argument precise is difficult because one wants to
recover exactly the probability distribution eq. (\ref{joint}). For
instance if some conversation are shorter than others, then they will
be accepted with higher probability, yielding a skewed
distribution. Nevertheless the above argument is very suggestive. For
instance in \cite{MBCC} it was shown that if the entangled state has
dimension $d$, then any measurement scenario 
 can be simulated in the average communication
model using less than $(6 + 3 \log_2 (d)) d + 2$ bits on
average. Combining this with the above argument suggests that if $\eta
< O(2^{-6d}d^{-3d})$ a local variable model should exist. This in turn
suggests that theorem 1 is close to optimal.

It is also interesting to combine the above argument with a result
from \cite{BCT} that states that it is always possible to simulate a
measurement scenario with $C = \log_2 |M_A|$ bits of
communication. Combining this with the above argument suggests that if 
$\eta > 1/|M_A|$ a local variable model should exist. This result
(in a slightly weaker form, since the result in \cite{BCT} 
depends only on $|M_A|$, independently of $|M_B|$) has
been proven by S. Popescu\cite{P} as follows :

{\em Theorem 3:} Consider a measurement scenario in which the number of
possible measurements is $|M_A|=|M_B|=M$. Then a local hidden variable 
model exists if the detector efficiency is $\eta= 1/M$.

{\em Proof:} The local hidden variable consists of the quadruple
$(x,i,y,j)$ where $x\in M_A$, $y\in M_B$, $i,j \in \{1,\ldots,d\}$ and $i,j$
have joint probabilities $P(i,j) = |\langle
\psi|x_i\rangle |y_j\rangle|^2$.
The protocol is as follows: Alice checks whether her measurement
is equal to $x$, if so she outputs $i$, if not she outputs $0$; Bob 
checks whether his measurement
is equal to $y$, if so he outputs $j$, if not he outputs $0$. 
This reproduces exactly the correlations eq. (\ref{joint}) with
$\eta = 1/M$.$\Box$

In summary we have presented a measurement scenario that closes the
detection loophole when the detector efficiency 
$\eta \simeq 2^{-0.0035 d}$ is exponentially small. 
This should be contrasted to the best previous result that required
$\eta > 2/3$ \cite{E}. Our measurement scenario 
requires an entangled system of large dimension $d$,
and it requires that Alice and Bob choose between 
exponentially 
many measurements. We have argued
that it is not possible to substantially improve this
measurement scenario, either by decreasing the number of measurements, 
or by decreasing the dimension, while keeping the same resistance to
inefficient detectors. 

The results reported here are inspired by recent work in 
 communication complexity. Indeed the measurement scenario
we consider in our main theorem is also 
known to require a large amount of communication in order to be
simulated classically\cite{BCT}, 
and our general 
arguments concerning bounds 
on the minimum detector efficiency required to close the
detection loophole follow from mappings between communication models
and local variable models with inefficient detectors. 
This connection between two different approaches to
entanglement, namely the point of view of computer scientists and the
more pragmatic considerations of experimentalists will, we hope,
continue to prove fruitful.

I would like to thank Nicolas Cerf, Richard Cleve, Thomas
Durt, Nicolas Gisin, Jan-Ake Larsson, Noah Linden and Sandu Popescu 
for helpful
discussions. 
Funding by the European Union under project EQUIP
(IST-FET program) is gratefully acknowledged.

\end{multicols}
\end{document}